%% file: main.tex
\documentclass[nonacm,sigplan,screen]{acmart}

\usepackage{adjustbox}
\usepackage{amsthm}
\usepackage{algorithm}
\usepackage[noend]{algpseudocode}
\usepackage{colortbl}
\usepackage{graphicx}
\usepackage{mathtools}
\usepackage{pifont}
\usepackage{threeparttable}
\usepackage{multirow}
\usepackage{multicol}
\usepackage{breakurl}
\usepackage{soul}
\usepackage{enumitem}
\setlist[itemize]{leftmargin=*}

\input{config}

\begin{document}

\title{Democratizing MoE inference on commodity GPUs with \sys{}}

\author{Ruwen Fan}
\affiliation{\institution{Tsinghua University}\country{}}
\email{frw23@mails.tsinghua.edu.cn}

\author{Yuezhi Zu}
\affiliation{\institution{Tsinghua University}\country{}}
\email{zuyz23@mails.tsinghua.edu.cn}

\author{Minhui Xie}
\affiliation{\institution{Nanjing University}\country{}}
\email{xieminhui@nju.edu.cn}

\author{Junru Li}
\affiliation{\institution{Alibaba Cloud Computing}\country{}}
\email{rusuo.ljr@alibaba-inc.com}

\author{Qingda Hu}
\affiliation{\institution{Alibaba Cloud Computing}\country{}}
\email{qingda.hqd@alibaba-inc.com}

\author{Xinjun (Jimmy) Yang}
\affiliation{\institution{Alibaba Cloud Computing}\country{}}
\email{xinjun.y@alibaba-inc.com}

\author{Jiwu Shu}
\affiliation{\institution{Tsinghua University}\country{}}
\email{shujw@tsinghua.edu.cn}

\author{Youyou Lu}
\authornote{Corresponding author.}
\affiliation{\institution{Tsinghua University}\country{}}
\email{luyouyou@tsinghua.edu.cn}

\begin{abstract}

Deploying Mixture-of-Experts (MoE) models relies heavily on Expert Parallelism, which generates intense inter-GPU communication. 
Consequently, state-of-the-art inference systems require high-bandwidth, P2P interconnects (e.g., NVLink) in datacenter GPUs to handle massive token routing, making deployment prohibitively expensive.
Consumer GPUs offer comparable compute power at significantly lower cost,
promising to democratize MoE inference for individuals and enable privacy-preserving local deployments.
However, their bandwidth-limited (only weak PCIe bus bandwidth) and host-mediated interconnects (no P2P support) introduce severe communication bottlenecks.

We present \sys{}, a communication-efficient MoE inference system that resolves this mismatch through novel \textit{host-centric routing}.
Our key insight is that the unique communication topology provides the opportunity to elevate the host to an active routing hub, which can fundamentally reduce communication volume and eliminate global synchronization-induced stalls.
Specifically,
for token dispatch, we introduce host-backed token multicast to write shared tokens to the host exactly once, eliminating outbound transmission redundancy.
For token combine, we propose a fine-grained, token-level aggregation mechanism using host staging buffers,
which replaces rigid global synchronization and mitigates straggler effects. 
Evaluation on RTX 5090 GPUs shows that \sys{} improves inference throughput by up to 1.46$\times$, 
approaching the performance of NVLink-capable A800 GPUs at only 23.4\% of the hardware cost.
\end{abstract}

\maketitle
\pagestyle{plain}

\input{content/2.intro}
\input{content/3.background}

\input{content/4.design}

\input{content/5.evaluation}

\input{content/6.relatedworks}
\input{content/7.conclusion}

\bibliographystyle{plain}
\bibliography{references}

\end{document}

%% file: config.tex
\newcommand{\sys}{\textsc{CoMoE}}

\usepackage{booktabs}

\newcommand{\parabf}[1]{\noindent\textbf{#1}}

%% file: content/2.intro.tex
\section{Introduction} 
Mixture-of-Experts (MoE) has emerged as a powerful paradigm for scaling LLMs by increasing parameter count without proportionally increasing computational cost~\cite{shazeer2017moe, gshard, switchtransformer, deepseekmoe}. 
Unlike dense architectures, MoE introduces a set of \textit{expert} subnetworks and a lightweight routing mechanism that dynamically selects a subset of experts for each input token. 
As modern MoE models trend toward fine-grained specialization with dozens or hundreds of experts,
their huge memory footprints necessitate multi-GPU deployments via Expert Parallelism (EP)~\cite{gshard, switchtransformer, deepspeed}.

\begin{figure}[t]
\centering
\includegraphics[width=\linewidth]{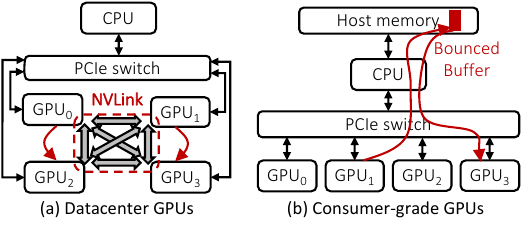}
\caption{Comparison of communication modes between datacenter and consumer-grade GPUs.}
\label{fig:intro:consumer_vs_datacenter}
\end{figure}

Deploying MoE models with EP inherently introduces a severe inter-GPU communication bottleneck: the token dispatch phase (routing tokens to distributed experts) and the combine phase (aggregating expert outputs) necessitate massive All-to-All collective communications
~\cite{aurora, deepspeed}.
Consequently,
state-of-the-art MoE inference systems~\cite{fastermoe, lina, deepspeed, tensorrt} rely on  
high-bandwidth, peer-to-peer (P2P) interconnects such as NVLink/NVSwitch on datacenter GPUs (e.g., A100/H100, Figure~\hyperref[fig:intro:consumer_vs_datacenter]{\ref*{fig:intro:consumer_vs_datacenter}(a)}).
Unfortunately, the high cost of these datacenter-grade ecosystems places MoE inference far beyond the budget of most individuals and organizations, creating a pressing need for more cost-effective solutions.

In contrast, consumer-grade GPUs (e.g., NVIDIA RTX 5090) offer an attractive alternative.
A flagship consumer GPU such as the RTX 5090 delivers raw computational power (FLOPs) and memory bandwidth on par with datacenter-grade GPUs like NVIDIA A800, while costing merely one-sixth of the datacenter counterpart.
This dramatic cost efficiency makes consumer GPUs a promising substrate to democratize MoE inference~\cite{moe-lightning, moe-infinity, mobile, eliseev2023fastmoeoffload}, as well as an ideal platform for privacy-preserving local deployments of intelligent applications~\cite{openclaw, privategpt, powerinfer, llamacpp}.

However, directly deploying existing MoE inference systems on consumer GPUs yields poor performance (as low as 61\% of the throughput compared to datacenter GPUs with similar FLOPs).
As our benchmarks reveal, this performance collapse is due to the dominant inter-GPU communication cost; on RTX 5090, communication overhead accounts for 81.6\% of the total execution time.
We identify that the root cause of this communication bottleneck is two-fold.
The first is an inherent hardware limitation.
Unlike datacenter GPUs, consumer GPUs connect via standard PCIe buses and lack P2P support (Figure~\hyperref[fig:intro:consumer_vs_datacenter]{\ref*{fig:intro:consumer_vs_datacenter}(b)}).
All cross-GPU traffic must be host-mediated (bounced through host DRAM), with a nominal bandwidth of only about one-sixth of NVLink's, and even less in practice.
Second, and more crucially, existing systems are fundamentally designed under the assumption of datacenter GPUs, which amplifies the hardware limitation.
In the absence of P2P, they fall back on All-to-All collective communication primitives to implement MoE routing.
However, such collective primitives are ill-suited for consumer GPU hardware.
They not only induce redundant transmissions from the source GPU to the host during dispatch (\S\ref{sec:mot:overhead1}, Analysis \#1),
but also enforce a rigid global barrier during combine, exposing communication latency on the critical path (\S\ref{sec:mot:overhead2}, Analysis \#2).

This paper presents \sys{}, a communication-efficient MoE inference system designed to unlock consumer-grade GPUs' full potential.
Our key insight is that the host-mediated, centralized communication topology of consumer GPU servers---long treated as an inconvenient limitation---can instead be turned into a powerful architectural advantage.
\textit{Rather than treating host memory as a passive bounce buffer, 
we explicitly elevate the host into an active, centralized routing hub}, and
rethink MoE's dispatch and combine phases around it to eliminate redundant data movement and synchronization stalls.

During the \emph{dispatch phase}, \sys{} adopts a host-backed token multicast to eliminate redundancy during token dispatch. 
Unlike traditional systems that view the dispatch phase of MoE from an All-to-All perspective, \sys{} adopts a token-centric perspective and treats token routing as a multicast problem.
Specifically,
\sys{} uses a centralized host buffer for token distribution.
A sender GPU writes each shared token to this buffer exactly once, accompanied by lightweight metadata, and destination GPUs then pull their required tokens in parallel.
This eliminates outbound data redundancy and shifts the bandwidth pressure from the bottlenecked sender to the aggregate inbound bandwidth of the receiving GPUs.

During the \emph{combine phase},
\sys{} proposes a fine-grained combine via host staging. 
Specifically,
to eliminate synchronous stalls of All-to-All primitives during the combine phase, 
\sys{} decomposes the global collective into one-sided, token-level memory operations using host memory as a staging area.
This transforms bulk-synchronous collectives into a completion-driven pipeline, overlapping data transfers with computation to fully mask straggler stalls. 
Fast GPUs can continuously deposit and drain ready tokens without waiting for slower ones.

We evaluate \sys{} using representative MoE models with the ShareGPT dataset on both RTX 5090 and A800 platforms. The results show that on the RTX 5090, \sys{} improves inference throughput by up to 1.46$\times$ over SGLang. Moreover, it brings TTFT close to that of an A800 system equipped with high-speed NVLink interconnects and reaches 85--90\% of the throughput of the strongest A800 configuration (SGLang with DeepEP). To deliver the same serving throughput as that configuration, \sys{} requires at most 23.4\% of the equivalent GPU investment. Beyond raw serving metrics, in real-world scenarios, for complex agentic workflows like DeepResearch, \sys{} consistently accelerates overall task execution, yielding a 13.1\% reduction in median latency.

Overall, we make the following contributions:
\begin{itemize}
    \item We identify the key bottleneck of MoE inference on consumer GPUs:
    limited inter-GPU interconnection severely increases communication overhead, and its cost is further amplified by the ill-suited collective communication primitives.
    \item We introduce \sys{}, a host-centric MoE communication system for consumer GPUs, featuring a host-backed token multicast for dispatch and a fine-grained host staging combine.
    \item We implement \sys{}, and the evaluation shows that it improves inference efficiency on consumer GPUs, delivering performance comparable to datacenter-grade GPUs at merely 23.4\% of the cost.
\end{itemize}

%% file: content/3.background.tex
\section{Background and Motivation}

\subsection{MoE Inference with Expert Parallelism}
Mixture-of-Experts (MoE) has become the de facto architecture for large language models (LLM), which effectively scales model capacity without a proportional increase in computational cost~\cite{shazeer2017moe, gshard, switchtransformer, deepseekmoe}. 
This is achieved through sparse activation: a router directs each input token to a small subset (Top-$k$) of \textit{experts} (subnetworks)~\cite{shazeer2017moe, gshard}.
The recent trend of modern MoE towards fine-grained specialization has driven the total expert count significantly higher—typically ranging from 64 to 128 or more (e.g., DeepSeek-v2~\cite{deepseek-v2}, Qwen3~\cite{qwen3}, and GLM-4.7~\cite{glm-4.7}).
Because the aggregated memory footprint of these dozens or hundreds of experts typically exceeds single-GPU capacity, 
this necessitates multi-GPU deployments via Expert Parallelism (EP), which distributes experts across multiple devices.

\begin{figure}[t]
\centering
\includegraphics[width=\linewidth]{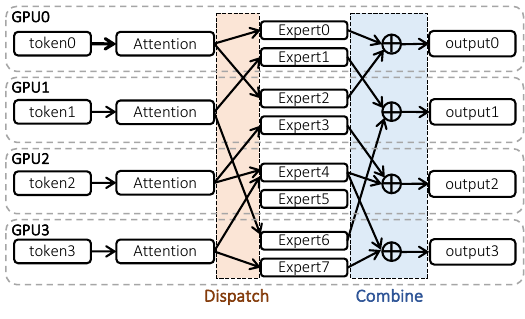}
\caption{Example of token routing in Expert Parallelism.} 
\label{fig:moe-ep-routing}
\end{figure}

This distribution, coupled with sparse routing, fundamentally alters the model's execution paradigm. Unlike dense layers where tokens follow a uniform, device-local computation path, MoE layers under EP must route tokens to experts scattered across different GPUs.
As illustrated in Figure~\ref{fig:moe-ep-routing}, 
each MoE layer imposes two critical communication phases: 
a \textit{dispatch} phase that sends tokens to their target GPUs, and a subsequent \textit{combine} phase that gathers the expert outputs to reconstruct the hidden states. 
Because routing decisions are input-dependent and tokens often target multiple experts, the resulting cross-GPU communication is highly irregular and unpredictable. 
Consequently, the overall efficiency of MoE inference is inextricably linked to the performance of the underlying inter-GPU communication substrate.

\begin{table}[t]
\centering
\caption{Comparison of datacenter GPUs and consumer-grade GPUs. Consumer GPUs offer comparable compute power but limited interconnect bandwidth.}
\label{tab:gpu_comparison}
\small
\begin{tabular}{lccc}
\toprule
\textbf{Specification} & \textbf{A800-SXM} & \textbf{RTX 5090} & \textbf{Ratio} \\
\midrule
FP16 TFLOPs & 312 & 419 & 1.34$\times$ \\
BF16 TFLOPs & 312 & 209.5 & 0.67$\times$ \\
Memory (GB) & 80 & 32 & 0.4$\times$ \\
Memory BW (TB/s) & 2.0 & 1.8 & 0.9$\times$ \\
Interconnect & NVLink & PCIe 5.0 & -- \\
Inter-GPU BW (GB/s) & 400 & 64$^\dagger$ & 0.16$\times$ \\
Price (USD) & \$12,000 & \$1,999 & 0.17$\times$ \\
\midrule
Cost/TFLOPs (FP16) & \$38.5 & \$4.8 & 0.12$\times$ \\
\bottomrule
\end{tabular}
\begin{flushleft}
\small $^\dagger$ PCIe 5.0 x16 bandwidth; effective bandwidth for inter-GPU communication is even lower due to host mediation.
\end{flushleft}
\end{table}

\subsection{Consumer-Grade GPUs}
Consumer-grade GPUs offer a compelling value proposition for democratizing MoE inference.
Beyond strong compute capability and memory bandwidth at much lower cost than datacenter GPUs, their affordability makes privacy-preserving, on-premise deployments viable.
These advantages render consumer GPUs uniquely attractive for the private deployment of intelligent assistant applications (e.g., OpenClaw~\cite{openclaw}), empowering individuals, small enterprises, and research laboratories to process sensitive data locally.
As detailed in Table~\ref{tab:gpu_comparison}, an NVIDIA RTX 5090 delivers memory bandwidth comparable to an A800 and tensor throughput of the same order (higher in FP16, lower in BF16), at roughly 17\% of the price; even in BF16, its cost per TFLOP is only a quarter of the A800's.

However, consumer GPUs suffer from a dramatically weaker communication substrate.
Figure~\ref{fig:intro:consumer_vs_datacenter} compares the communication schemes of the two GPU types.
On datacenter GPUs, high-bandwidth, point-to-point interconnects like NVLink and NVSwitch provide extremely high communication bandwidth (e.g., 400 GB/s) and direct memory load/store capability between GPUs (e.g., via NVIDIA NVSHMEM~\cite{nvshmem}).
In contrast,
consumer GPUs lack support for PCIe peer-to-peer (P2P) communication in multi-GPU deployments~\cite{li2020evaluating, deepspeed, huang2023towardsmoedeployment, megascale}.
Every cross-GPU transfer must traverse the PCIe fabric and bounce through host memory; these transfers therefore compete for a much narrower communication substrate and additionally consume host-side bandwidth and synchronization resources.

This communication gap is particularly damaging for MoE workloads.
Unlike dense models that primarily rely on predictable, coarse-grained parameter synchronization,
MoE inference under EP inherently generates irregular, fine-grained data movement.
Each MoE layer triggers token dispatch and combine phases, where the payloads are small (token-level) and the routing destinations change dynamically from batch to batch.
On datacenter systems, the immense bandwidth and P2P load/store of NVLink largely mask the overhead of this irregular traffic.
On consumer hardware, however, funneling this granular data exchange through a host-mediated PCIe bottleneck stalls the GPU execution pipeline.
Ultimately, this mismatch shifts the system bottleneck from expert computation to interconnect efficiency, threatening to negate the raw computing advantages of consumer GPUs.

Several prior works have explored communication optimization for MoE inference
under EP, most notably DeepSeek DeepEP~\cite{deepep} and PPLX-kernels~\cite{pplx}.
While these systems achieve strong performance on datacenter hardware, they are fundamentally predicated on hardware capabilities absent from consumer platforms, making direct adoption infeasible.

\subsection{Motivation}

\begin{figure}[t]
\centering
\includegraphics[width=\linewidth]{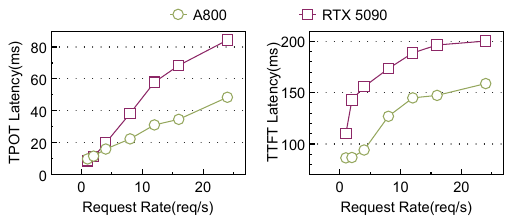}
\caption{Time to first token (TTFT) and time per output token (TPOT) comparison of SGLang on 8 NVIDIA A800 (datacenter) and RTX 5090 (consumer) GPUs.}
\label{fig:moti}
\end{figure}

\begin{figure}[t]
\centering
\includegraphics[width=\linewidth]{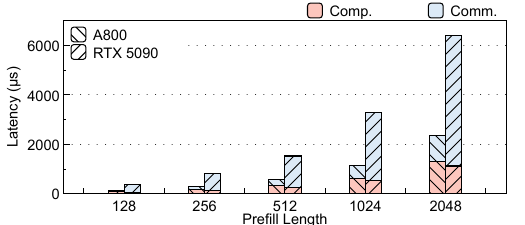}
\caption{Time breakdown of a prefill iteration. Comm: communication; Comp: computation.
}
\label{fig:moti:break}
\end{figure}

\parabf{Experiment: Impact of consumer GPUs on MoE model inference.} To quantify the resulting performance gap, we take SGLang~\cite{sglang} as a representative system and conduct a microbenchmark on both datacenter and consumer GPU platforms. 
As shown in Figure~\ref{fig:moti}, latency on consumer GPUs increases by up to 1.97$\times$ compared to datacenter GPUs. To understand the root cause of this gap, we decompose the model execution time into two components: collective communication (\textit{comm.}) and computation (\textit{comp.}), with detailed configurations described in \S\ref{eva:tech}. Figure~\ref{fig:moti:break} reveals that the performance gap is almost entirely attributable to the communication component, whereas computation time remains comparable across platforms. When the prefill length is 2,048, communication accounts for 81.6\% of the total execution time on the RTX 5090 GPU, making it the dominant contributor to inference latency. In contrast, on datacenter GPUs, this proportion is only 44.8\%.

\parabf{Analysis.} We identify that the root cause of the elevated communication overhead is the stark limitation in collective communication bandwidth on consumer GPUs, which rely on PCIe interconnects rather than high-speed NVLink. However, raw bandwidth scarcity alone does not fully account for the severity of the performance collapse. Rather, the impact of this bandwidth deficit is severely amplified by a structural mismatch between the dynamic characteristics of MoE routing and the rigid semantics of traditional collective primitives (\textit{e.g.}, NCCL All-to-All). This amplification manifests in two critical dimensions:

\begin{figure}[t]
\centering
\includegraphics[width=\linewidth]{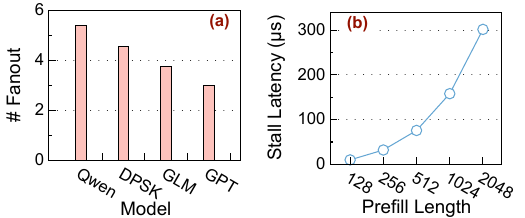}
\caption{(a) Average token-to-GPU dispatch fanout across models. See detailed model configuration in Table~\ref{tab:eva:model}. (b) Stall latency caused by coarse-grained synchronization.
}
\label{fig:moti:analysis}
\end{figure}

\ul{\textit{1) Data redundancy in dispatch exacerbates bandwidth starvation.}} 
During the dispatch phase, MoE routing frequently assigns a single token to multiple experts. 
When these experts reside on different GPUs, the token must be multicast to different destinations. As 
Figure~\hyperref[fig:moti:analysis]{\ref*{fig:moti:analysis}(a)}
shows, the average token is routed to 2.98 to 5.37 distinct GPUs depending on the model architecture. For instance, in the Qwen3-30B-A3B model, which activates 8 experts per token, a single token must be transmitted to an average of 5.37 different GPUs.
Standard All-to-All interfaces are completely oblivious to token-level redundancy, forcing the source GPU to transmit duplicate copies of the exact same token to different destinations. 
On datacenter GPUs, the immense bandwidth of NVLink masks this inefficiency. 
On consumer GPUs, however, multiplying the dispatch traffic over already-starved PCIe links severely exacerbates egress contention, directly amplifying the bandwidth penalty.\label{sec:mot:overhead1}

\ul{\textit{2) Coarse-grained synchronization in combine prevents latency hiding.}} 
The dynamic nature of MoE routing inherently creates load imbalance (the ``straggler'' effect). 
Standard All-to-All collectives are bulk operations: a GPU's transfers with a peer start only once that peer enters the collective, and the subsequent reduction cannot start until the slowest peer's data has arrived. As Figure~\hyperref[fig:moti:analysis]{\ref*{fig:moti:analysis}(b)} shows, this stall time increases with the prefill length; when the prefill length is 2,048, it can reach 301\textmu s, accounting for up to 11.9\% of the combine phase.
Consequently, fast GPUs are forced to stall and wait for laggards, completely wasting valuable opportunities to overlap computation with communication. State-of-the-art datacenter solutions (\textit{e.g.}, DeepEP~\cite{deepep}) circumvent this by exploiting NVLink's hardware-level peer-to-peer (P2P) memory access for asynchronous transfers. 
Stripped of such P2P capabilities, consumer systems are forced back into synchronous, coarse-grained primitives.
Because communication cannot be effectively overlapped, the full latency penalty of the low-bandwidth PCIe links is starkly exposed on the critical path, further compounding the performance degradation.\label{sec:mot:overhead2}

These observations motivate a communication design specialized for consumer GPU servers. Rather than adapting primitives built for NVLink-class interconnects, we need mechanisms that explicitly account for host-mediated communication, centralized PCIe topology, and the scarcity of inter-GPU bandwidth.

%% file: content/4.design.tex
\section{Design and Implementation}
\begin{figure*}[t]
\centering
\includegraphics[width=\linewidth]{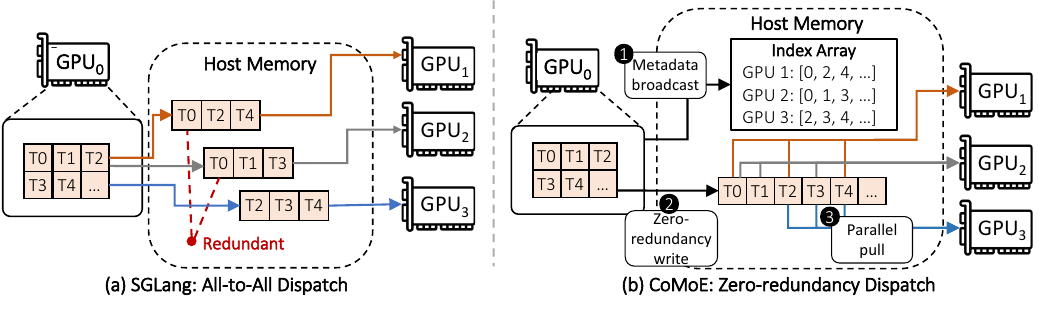}
\caption{
Comparison of two dispatch schemes.
\textit{(a) Existing system SGLang~\cite{sglang} exhibits transmission redundancy when dispatching tokens via All-to-All primitives.
(b) \sys{} achieves zero-redundancy via host-backed token multicast.
}
} 
\label{fig:dispatch}
\end{figure*}

We introduce \sys{}, a communication-efficient MoE inference system tailored for consumer GPUs, to fully leverage the cost-effectiveness advantages of consumer hardware. 
Our key insight is that the host-mediated, centralized communication topology inherent to consumer GPU servers, long dismissed as a cumbersome drawback, can be repurposed into a compelling architectural strength.

\sys{} embraces the host-mediated nature of consumer PCIe servers and turns it into a first-class design principle.
Rather than treating host memory as a passive bounce buffer, we explicitly elevate the host into an active, centralized routing hub.
Specifically,
\sys{} introduces \ul{\textit{zero-redundancy dispatch via host-backed token multicast}}~(\S\ref{sec:host_centric}), which recasts the MoE dispatch phase as a token-level multicast, exploiting the host memory and PCIe topology to eliminate redundant data transfers. 
Furthermore, \sys{} employs a \ul{\textit{fine-grained combine via host staging}}~(\S\ref{sec:combine}), which turns the host into an asynchronous rendezvous layer, which enables token-level pipelining of transfer and aggregation and creates more opportunities to overlap communication with aggregation.

\subsection{Zero-Redundancy Dispatch via Host-Backed Token Multicast}
\label{sec:host_centric}

We begin by theoretically analyzing the communication redundancy of existing All-to-All interfaces for token dispatch, and subsequently propose our solution based on host-backed token multicast.

\parabf{Transmission redundancy in token dispatch with conventional All-to-All primitives.}
Conventionally, the dispatch phase is modeled as an All-to-All collective. While this abstraction loosely captures the communication pattern, it mischaracterizes the data semantics.
Unlike a true All-to-All primitive, where every message is unique to its destination, MoE routes the same token to multiple experts that may reside on different GPUs~(Figure~\hyperref[fig:dispatch]{\ref*{fig:dispatch}(a)}).
Modern MoE models typically activate $k \in \{4, 8\}$ experts per token out of
$E \in \{64, 128\}$ total experts. Under expert parallelism across $G = 8$ GPUs,
each GPU hosts $E/G$ experts. Assuming the $k$ experts are selected uniformly without replacement,
the expected number of distinct destination GPUs of a single token is:
$$ \mathbb{E}[N_{\text{dst}}] = G \left( 1 - \frac{\binom{E - E/G}{k}}{\binom{E}{k}} \right)$$
which yields approximately \textbf{3.38 GPUs} for $E=64, k=4$, and \textbf{5.34 GPUs} for $E=128, k=8$. 
This means that under the conventional All-to-All formulation, the source GPU redundantly transmits 
the same token to roughly 3 to 5 of the 8 GPUs on average.
Treating this shared data delivery as All-to-All forces the source GPU to transmit redundant copies of the exact same token (one per destination). 
This inflates source-side egress traffic in proportion to the number of remote destinations per token, creating a bottleneck on bandwidth-constrained PCIe links.

\parabf{Rethinking MoE dispatch: A new token-level multicast perspective.} 
Our core observation is that, from the perspective of an individual token, MoE dispatch is fundamentally a fine-grained, one-to-many multicast. 
In an ideal multicast scenario, a source GPU should only transmit a token exactly once, regardless of its fanout. 
However, this one-to-many software semantic fails to propagate down to the actual physical routing paths. 
Constrained by the strict point-to-point nature of standard PCIe links, the multicast intent inevitably degenerates into redundant transmissions under the hood.
To physically realize zero-redundancy dispatch, 
we draw on the mature multicast paradigm from traditional networking: a centralized ``switching hub'' is required to handle data replication for one-to-many delivery, rather than forcing the sender to transmit redundant data copies.

We observe that consumer GPU servers inherently expose an analogous structure: all GPUs are connected to the host (CPU and DRAM) via PCIe. This centrally accessible host memory can be repurposed as a native multicast hub. 
By shifting the paradigm from point-to-point communication to a shared-hub architecture, a token only needs to leave the source GPU exactly once. Instead of duplicating data across slow PCIe interconnects, the source GPU writes the token to a shared buffer in host memory, where it instantly becomes readable by all GPUs on the fabric. This perfectly mirrors multicast semantics and completely bypasses point-to-point redundancies.

Capitalizing on this architectural observation, \sys{} introduces \textit{Host-backed Token Multicast}, a mechanism that realizes token-level multicast over the PCIe topology without any specialized networking hardware.
As depicted in Figure~\hyperref[fig:dispatch]{\ref*{fig:dispatch}(b)}, \textit{host-backed token multicast} contains three main steps.
\ding{182}~\textbf{Metadata broadcast.} Based on the rank of the expert activated by each token, the sender collects the token data that needs to be received for each rank to form an index array. 
The sender GPU writes a per-target index array that specifies which tokens each recipient GPU is responsible for fetching.
\ding{183}~\textbf{Zero-redundancy write.} The sender categorizes all tokens into two types: those computed locally and those to be sent. Each token to be sent is written directly into the host memory \emph{exactly once}. 
\ding{184}~\textbf{Parallel pull.} Guided by the index array, each destination GPU independently pulls ready tokens from host memory.
Since all receivers fetch from the same host-resident copy in parallel, multicast load is shifted from redundant source-side replication to the aggregate host-to-GPU read bandwidth of the receivers.

As the source GPU issues only a single write per token, rather than one write per (token, destination) pair, the egress pressure on the source GPU's PCIe uplink is reduced proportionally to the average token-to-GPU fanout $\mathbb{E}[N_{\text{dst}}]$, directly alleviating the outbound bandwidth bottleneck that plagues high-fanout MoE configurations. 
The indirection through host memory transforms what would otherwise be a one-to-many push into a many-from-one pull. \sys{} writes each remote token to host memory once and allows all required destination GPUs to fetch it independently, thereby eliminating the source-side bandwidth amplification caused by repeated token transmissions over the PCIe link. This removes the source-side bandwidth amplification caused by token fanout and shifts the remaining communication demand to the aggregate capacity of the shared PCIe and host-memory subsystem.

\subsection{Decomposed Fine-grained Combine via Host Staging}\label{sec:combine}

\begin{figure}[t]
\centering
\includegraphics[width=\linewidth]{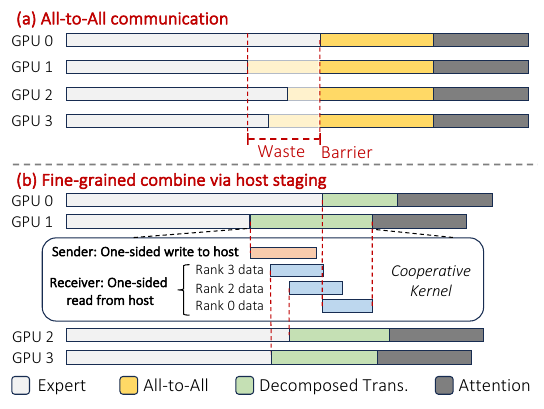}
\caption{Comparison of traditional All-to-All and \sys{}'s decomposed combine.}
\label{fig:des:combine}
\end{figure}

In this section,
we begin with a foundational analysis that upends a core assumption of existing MoE systems: the synchronous All-to-All primitives universally relied on for the combine phase are not an inherent, mandatory requirement.
Then, we introduce a decomposed combine protocol mediated by host-side staging buffers, decoupling senders from receivers and leveraging the host memory as an active stash.
Beyond cross-GPU data transfer, the combine phase requires a token-wise reduction to aggregate outputs from multiple experts for each input token. 
We detail our co-design of this reduction computation with the decomposed transfer pipeline.
Finally, we introduce two optimizations: tiled transfer and batched commit.

\parabf{Synchronous All-to-All primitives are not necessary.}
Traditional synchronous All-to-All primitives are highly vulnerable to MoE's dynamic load imbalance~(Figure~\hyperref[fig:des:combine]{\ref*{fig:des:combine}(a)}); they stall global transmission for stragglers and block arithmetic reduction until all data arrives, squandering opportunities to overlap communication with computation. 

Actually, the reconstruction of a specific token $i$  only requires the arrival of its assigned expert contributions. 
It does not, and should not depend on the completion of unrelated tokens or the progress of other GPUs. 
Thus, the combine operations of MoE do \emph{not} require global collective completion. 
The minimal synchronization unit for MoE is the token, not the batch.

\parabf{Decomposing All-to-All into one-sided transfers.} 
To exploit this, we observe that the host memory in consumer GPU servers can be promoted from a passive bounce buffer to an active staging stash.
By decomposing the All-to-All collective into independent, one-sided writes (producer) and reads (consumer),
\sys{} introduces a fine-grained decomposed combine mechanism.
Thus,
we can decouple the progress of the sender GPUs from the receivers and bypass these synchronization bottlenecks. 
Guided by this insight, \sys{} redesigns combine as a two-phase communication path composed of sender-side deposits and receiver-side drains, coordinated only through host-resident ready flags and per-token completion state.

\begin{figure}[t]
\centering
\includegraphics[width=\linewidth]{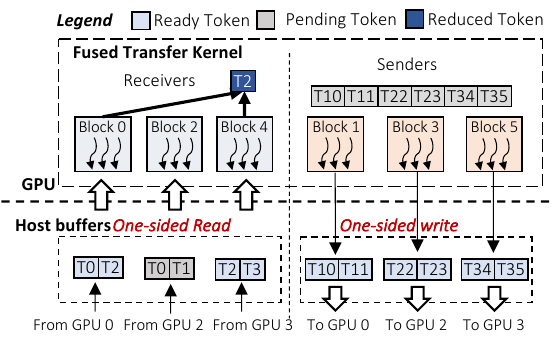}
\caption{Fused transfer kernel for the combine phase.
\textit{Assume this kernel runs on GPU 1: sender blocks 1/3/5 send tokens to the corresponding staging buffers on GPUs 0/2/3, and receiver blocks 0/2/4 read from the staging buffers of GPUs 0/2/3.}
}
\label{fig:des:combine_kernels}
\end{figure}

To implement this at line-rate PCIe speeds,
\sys{} tailors a specialized fused transfer kernel, as illustrated in Figure~\ref{fig:des:combine_kernels}.
\sys{} specializes GPU thread blocks into two roles, sender and receiver, and fuses them with the cooperative kernel technique (i.e., \texttt{cudaLaunchCooperativeKernel}) to saturate the bidirectional read and write bandwidth of PCIe. 
The data transfer for each token contains two dependent, one-sided operations (a one-sided write and a one-sided read), with no inter-token dependencies whatsoever.
Sender blocks continuously deposit locally computed expert outputs into host stash buffers and publish ready flags once a stash slot becomes valid. 
Receiver blocks independently poll these stash buffers, drain any ready contribution into temporary device buffers, and update the completion state of the corresponding token. 
Because deposits and drains are one-sided operations linked only through host-side readiness metadata, the two sides no longer need to advance in lockstep. 
A fast sender can keep depositing results even if the destination has not yet consumed earlier ones, while a fast receiver can keep draining any already-ready contribution without waiting for slower senders that are working on unrelated tokens. 
This design is particularly important on consumer GPUs, where PCIe bandwidth is scarce and any unnecessary global synchronization directly exposes the laggard effect on the critical path.

\begin{figure}[t]
\centering
\includegraphics[width=\linewidth]{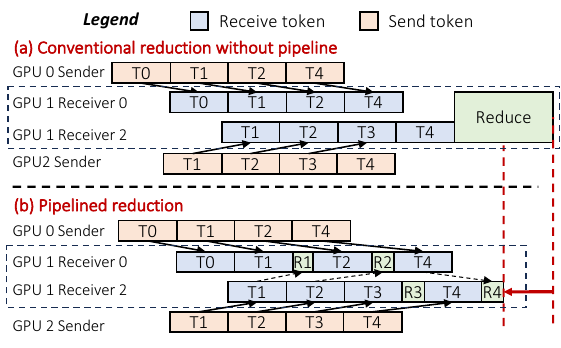}
\caption{Pipelined reduction.\textit{We assume that token $T_{0/1/2/4}$ processed during inference on GPU 1 are routed to experts on GPU 0, while $T_{1/2/3/4}$ are routed to experts on GPU 2.}}
\label{fig:des:pipelinedreduction}
\end{figure}

\parabf{Pipelined reduction.}
A conventional combine implementation includes two phases: 1) data transfer, all ranks' outputs are first transferred, and 2) only after the last transfer completes does each GPU launch a reduction kernel to aggregate (Figure~\hyperref[fig:des:pipelinedreduction]{\ref*{fig:des:pipelinedreduction}(a)}). 
The active staging buffer allows \sys{} to replace this with pipelined reduction (Figure~\hyperref[fig:des:pipelinedreduction]{\ref*{fig:des:pipelinedreduction}(b)}). 
For each token, \sys{} maintains a counter recording the number of outstanding rank-level contributions. Every time a receiver block drains one ready contribution from the host stash, it decrements the corresponding counter. 
The block that drives the counter to zero immediately invokes the final accumulation for that token, without waiting for a batch-wide barrier. 
This allows the tail of communication for one subset of tokens to be overlapped with useful combine computation on another subset, substantially reducing exposed latency in the presence of stragglers.

\parabf{NUMA-aware staging buffer placement.} 
For multi-GPU systems interconnected via PCIe, optimizing communication requires careful consideration of the underlying NUMA architecture.
In such topologies, the latency asymmetry between memory read and write operations across NUMA nodes becomes a critical bottleneck. A remote memory write over PCIe is inherently a unidirectional ``push'' operation that incurs minimal stalling. Conversely, a remote memory read is a ``pull'' operation requiring a full round-trip across the interconnect to issue a request and fetch the payload, making it significantly more latency-bound. To mitigate this, \sys{} deliberately forces the sending GPUs to directly write their computation results into staging buffers allocated within the receiving GPU's local NUMA memory domain. By anchoring the buffer to the receiver's side, senders can leverage efficient remote writes, while the receiver executes low-latency local reads. Furthermore, replacing round-trip remote reads with one-way remote writes reduces the PCIe transaction footprint (by eliminating read requests), which directly alleviates the inbound congestion and significantly reduces the overall end-to-end latency of the combine phase.

\parabf{Optimizations.}
We further introduce two optimizations to facilitate the decomposed combine.

The first is \textit{tiled transfer}. Because different models have varying intermediate state sizes (i.e., the per-token data size differs across models), transferring data at token granularity often fails to fully saturate PCIe bandwidth. If an entire block is used to transmit the data of a token, it may happen that the hidden states size of the model cannot be divided by the block size, resulting in SM idling during data transmission. Choosing a block size that is too small or a non-power of 2 will result in low SM utilization. To address this, we propose a Thread Tile Pipeline scheme. Specifically, we partition a single thread block into multiple thread tiles according to the per-token data size, where each thread tile is responsible for transferring one token.

The other is \textit{batched commit}.
The fine-grained transfer protocol introduced above creates a new challenge: if every token deposit triggers an independent synchronization action, signaling overhead scales linearly with the number of tokens, quickly becoming a bottleneck of its own. \sys{} therefore amortizes synchronization by allowing all thread tiles within a block to share commit state. Multiple token deposits or drains can be committed in batches, reducing the number of host-visible synchronization events while preserving token-level readiness. 
This design keeps the stash protocol lightweight and makes the decoupled combine path practical even when the number of transferred tokens is large.

\subsection{Implementation}\label{sec:impl}
We implement \sys{} as a new MoE communication backend for SGLang, alongside its existing All-to-All backends; the model code and scheduler are unchanged, and the backend is selected through SGLang's MoE backend configuration.
The staging buffers are ordinary host allocations that \sys{} pins and maps into every GPU's address space with \texttt{cudaHostRegister}. GPU kernels therefore access them directly with loads and stores over PCIe, and no \texttt{cudaMemcpy} or CPU thread lies on the data path.

\parabf{Host-side overhead.}
\sys{} adds no CPU computation. Although host memory serves as the routing hub, all transfers and synchronization are issued by the GPUs themselves; the CPU does not route, copy, or signal.
The main host-side cost is memory for the staging buffers. Each GPU's combine staging buffer holds up to $T_{\max}$ tokens from each of the $R$ GPUs, i.e., $T_{\max} \times R \times H \times \mathit{sizeof}(\mathit{dtype})$ bytes, where $T_{\max}$ is the maximum number of tokens per GPU in a batch and $H$ is the hidden size.
The dispatch phase reuses these buffers, so it requires no additional host memory; beyond them, \sys{} keeps only small metadata (index arrays and ready flags).
For Qwen3-30B-A3B ($H = 2{,}048$) with $R = 8$ and $T_{\max} = 16{,}384$, this is 512\,MiB per GPU, or 4\,GiB for all eight GPUs; for GPT-OSS-20B ($H = 2{,}880$), it is 720\,MiB per GPU (5.6\,GiB in total). Both are under 0.5\% of our server's 1.5\,TB of DRAM.

\parabf{SM usage.}
Because the transfer kernel moves data with SM loads and stores, it occupies SMs that would otherwise run computation. A sweep over the number of SMs shows that PCIe bandwidth saturates at about 24 SMs.
On the RTX 5090, \sys{} therefore launches 48 thread blocks of 512 threads; since each SM hosts two such blocks, the kernel occupies 24 of the 170 SMs (14.1\%).
This small footprint leaves most SMs free for concurrent computation, so \sys{} remains compatible with techniques that overlap communication with computation, such as two-batch overlap.

%% file: content/5.evaluation.tex
\section{Evaluation}

\subsection{Experimental Setup}

\parabf{Testbed.} 
Experiments are performed on a server with two Intel Xeon Platinum 8575C CPUs, 1.5TB DRAM, and 8$\times$ NVIDIA RTX 5090 GPUs. 
For cross-platform comparison, we also evaluate on a datacenter server with 8$\times$ A800 GPUs interconnected by NVLink (400 GB/s), equipped with two AMD EPYC 7742 CPUs and 2TB DRAM.
We use SGLang 0.5.9 and transformers 5.0.0.dev (a modified version to support GLM-4.7-Flash) for all the baselines. On the A800 platform, we additionally evaluate DeepEP v1.2.1 (patched for SM80 support) with NVSHMEM 3.3.9 for comparison.

\begin{table}[b] 
    \small
    \centering
    \caption{MoE models. Size: total parameters; Hid. Dim.: hidden size; Exp.: routed experts per MoE layer; Active Exp.: experts activated per token.}
    \label{tab:eva:model}
    \begin{tabular}{lllll}
        \toprule
        \textbf{Model}       & \textbf{Size} & \textbf{Hid. Dim.}  & \textbf{Exp.}  & \textbf{Active Exp.} \\ \midrule
        DeepSeek-V2-Lite~\cite{deepseek-v2}     & 16B       & 2048   & 64  & 2(S) + 6 \\         
        GPT-OSS-20B~\cite{gpt-oss}        & 20B       & 2880   & 32  & 4\\
        Qwen3-30B-A3B~\cite{qwen3}        & 30B       & 2048   & 128 & 8 \\         
        GLM-4.7-Flash~\cite{glm-4.7}        & 31B       & 2048   & 64  & 1(S) + 4\\
        \bottomrule
    \end{tabular}
    \begin{flushleft}
    (S): shared experts, always active.
    \end{flushleft}
\end{table}

\parabf{Models and Dataset.} 
We evaluate four representative MoE model configurations, as summarized in Table~\ref{tab:eva:model}. These models span diverse architectural designs and configurations, varying in model scale, the total number of experts, and the number of activated experts per token, thereby providing a representative coverage of modern MoE deployments. All models are evaluated in BF16 precision, following common practice. For benchmarking, we adopt the ShareGPT~\cite{sharegpt} dataset, which consists of real-world conversational data and is widely used to evaluate LLM inference performance under realistic workloads. Additionally, we evaluate \sys{} in realistic agent-based scenarios using the DeepResearch~\cite{tongyidr} agent and the BrowseComp-VL~\cite{browsecomp} dataset, in order to assess its performance under practical, end-to-end application workloads.

\parabf{Baseline.} We evaluate \sys{} against three primary baselines. (1) The SGLang default All-to-All backend on the 8$\times$RTX 5090 server, which uses NCCL for MoE communication. (2) The same serving stack on an 8$\times$A800 NVLink server using the default NCCL backend. By leveraging high-speed NVLink interconnects, this baseline provides a hardware-level reference point to assess how closely \sys{} on PCIe GPUs can approach datacenter-grade performance. (3) SGLang with the DeepEP\cite{deepep} backend on the 8$\times$A800 server, which represents a specialized MoE communication backend optimized for expert parallelism. We evaluate all three baselines for both throughput and latency.

\parabf{Metrics.}
We report the following primary metrics: (1) \emph{Time to First Token} (TTFT), which measures the latency from when a request is received to when the first output token is generated, capturing the prefill efficiency; (2) \emph{Time Per Output Token} (TPOT), which measures the average decoding latency per generated token; (3) \emph{Throughput} in tokens per second, measured as the total number of output tokens generated per unit time across all concurrent requests; and (4) \emph{End-to-end task latency} of real-world agentic tasks, including external tool calls and agent overhead. Where applicable, we also report MoE communication latency to provide insight into the communication overhead under different hardware and software configurations.

\subsection{Technique Analysis}\label{eva:tech}

\begin{figure}[t]
\centering
\includegraphics[width=\linewidth]{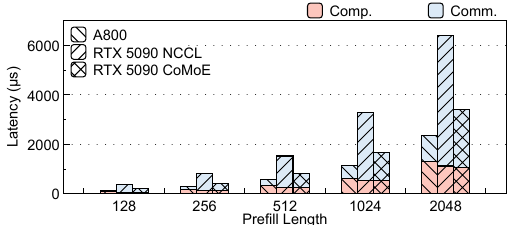}
\caption{Time breakdown of a prefill iteration.}
\label{fig:eva:breakdown}
\end{figure}

\parabf{Communication Time Percentage.} We evaluate communication efficiency using the Qwen3-30B-A3B model by synthetically generating inputs of lengths 128, 256, …, 2048 tokens, and profiling a single prefill iteration to measure the ratio of communication time to computation time. As shown in Figure~\ref{fig:eva:breakdown}, the baseline suffers from severe communication overhead, with the communication-to-compute ratio exceeding 4.6$\times$ across all tested sequence lengths — meaning communication dominates over 80\% of total prefill time. \sys{} reduces prefill communication latency by an average of 55.7\% compared to the baseline, bringing the ratio down to approximately 2.25$\times$, a 2.17$\times$ improvement. Furthermore, while the A800's datacenter-grade NVLink interconnect achieves a ratio of around 0.8$\times$, \sys{} significantly narrows the gap between consumer and datacenter hardware, demonstrating that our approach can effectively recover much of the interconnect disadvantage inherent to consumer GPUs.

\begin{figure}[t]
\centering
\includegraphics[width=\linewidth]{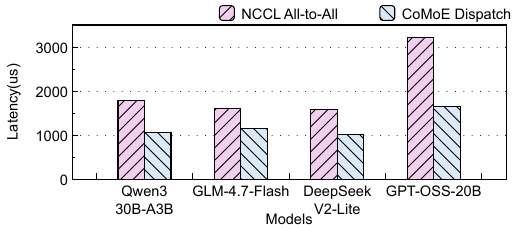}
\caption{Dispatch time comparison across different models.}
\label{fig:eva:redudency}
\end{figure}

\parabf{Dispatch Redundancy Elimination.} To verify that \sys{} eliminates the redundant transfer in dispatch, we traced the dispatch top-k data of 1,024 tokens across four models, and replayed them using \sys{} dispatch and NCCL All-to-All separately to benchmark their respective latencies in Figure~\ref{fig:eva:redudency}. Specifically, for Qwen3-30B-A3B, \sys{} reduces dispatch latency from 1789.7 µs to 1065.6 µs, yielding a speedup of 1.68$\times$. For GLM-4.7-Flash, DeepSeek-V2-Lite and GPT-OSS-20B, \sys{} achieves speedups of 1.40$\times$, 1.54$\times$ and 1.96$\times$, respectively. 

It is worth noting that GPT-OSS-20B has a substantially larger hidden state size of 2,880, which leads to significantly higher communication volume compared to the other models. As a result, the All-to-All baseline incurs considerably greater dispatch overhead for GPT-OSS-20B. For other models, the All-to-All baseline incurs lower latency on DeepSeek-V2-Lite (6 activated experts) and GLM-4.7-Flash (4 activated experts) than on Qwen3-30B-A3B (8 activated experts), since fewer activated experts implies fewer distinct destination GPUs and thus less data to be transmitted on average, which is consistent with the analysis in \S\ref{sec:host_centric}. In contrast, \sys{}'s Zero-Redundancy design ensures that each GPU transmits a fixed amount of data regardless of the number of activated experts—determined solely by the hidden state size, which is identical across other three models—resulting in nearly constant dispatch latency across all configurations.

\begin{figure}[t]
\centering
\includegraphics[width=\linewidth]{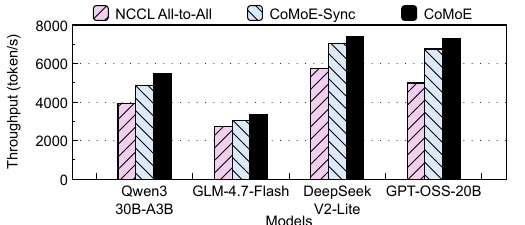}
\caption{Performance comparison of the combine phase with and without global synchronization.}
\label{fig:eva:combine_sync}
\end{figure}

\parabf{Impact of Host-Mediated Decoupled Combine.} To isolate and quantify the performance benefits of our decoupled design, we construct a controlled baseline by forcing \sys{} into a synchronized execution model~(\sys{}-Sync). As discussed in Section \ref{sec:combine}, traditional synchronous All-to-All primitives are highly vulnerable to MoE's dynamic load imbalance, which is especially noticeable under heavy loads. To emulate this limitation, we deliberately inject an artificial global barrier immediately after the expert computation phase within \sys{}. This barrier prevents any data transmission from starting until all GPUs have completed their locally imbalanced computations, forcing the system to behave like a conventional, rigid collective.

As Figure~\ref{fig:eva:combine_sync} shows, \sys{} consistently outperforms the synchronized baseline across all evaluated workloads. The largest gains appear on Qwen3-30B-A3B and GLM-4.7-Flash, where \sys{} improves throughput over \sys{}-Sync by 13.2\% (from 4,865.47 to 5,506.00) and 11.2\% (from 3,029.06 to 3,367.13), respectively. These gains indicate that both models suffer from noticeable dynamic load imbalance during the expert computation phase, causing the synchronous collective to stall for laggards. By employing host-mediated one-sided transfers, \sys{} effectively masks these straggler effects, allowing fast receivers to drain ready tokens eagerly. Furthermore, even for models with inherently higher baseline performance, such as DeepSeek-V2-Lite and GPT-OSS-20B, \sys{} still delivers 4.9\% and 8.1\% improvements, respectively. These empirical results validate that decoupling the sender and receiver progress is crucial for maximizing throughput and mitigating synchronization overheads in MoE combine phases.

\begin{figure}[t]
\centering
\includegraphics[width=\linewidth]{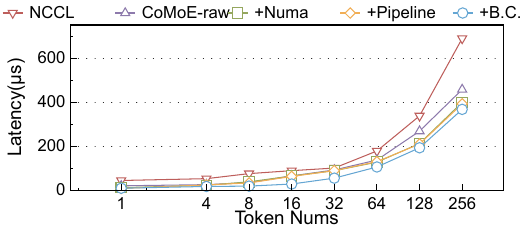}
\caption{\sys{} combine latency across ablations.}
\label{fig:eva:ablation}
\end{figure}

\parabf{Ablation of Combine.}
Using the hidden size (2048) and expert-per-token (6) configuration of DeepSeek V2 Lite, we generated random combine requests spanning 1 to 256 tokens and measured end-to-end combine latency (Figure~\ref{fig:eva:ablation}).
Removing global synchronization overhead (\sys{}-raw) already yields a $1.55\times$ geometric-mean speedup over baseline.
NUMA-aware buffer placement further reduces latency for larger transfers (269.4$\to$213.0 \textmu s at 128 tokens; 457.9$\to$399.7 \textmu s at 256 tokens), confirming that anchoring the stash buffer on the receiver side converts expensive remote reads into cheap local drains.
The thread-tile pipeline better saturates PCIe bandwidth at small-to-medium token counts, cutting 1-token latency from 13.3 to 9.5 \textmu s and 8-token latency from 39.9 to 34.8 \textmu s.
Batched commit delivers the largest incremental gain by amortizing per-token readiness signaling (64.2$\to$29.9 \textmu s at 16 tokens; 90.5$\to$56.5 \textmu s at 32 tokens).
Overall, the full design achieves a $2.46\times$ geometric-mean speedup over NCCL All-to-All and a $1.59\times$ speedup over \sys{}-raw, reducing average latency by $48.9\%$ and $27.5\%$, respectively, demonstrating that \sys{}'s gains stem from both decomposing All-to-All into one-sided transfers and systematically eliminating the secondary bottlenecks of host-mediated communication.

\subsection{Overall performance}

\begin{figure*}[t]
\centering
\includegraphics[width=\linewidth]{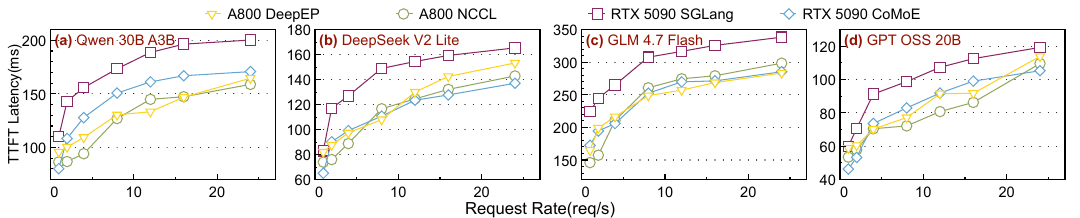}
\caption{TTFT across different request rates.}
\label{fig:eva:ttft}
\end{figure*}

\begin{figure*}[t]
\centering
\includegraphics[width=\linewidth]{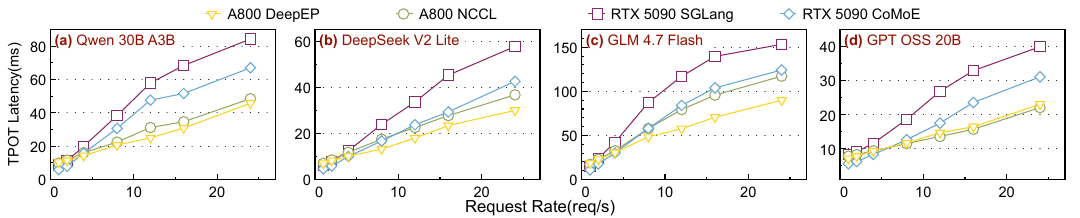}
\caption{TPOT across different request rates.}
\label{fig:eva:tpot}
\end{figure*}

\parabf{Latency.} Figure~\ref{fig:eva:ttft} and Figure~\ref{fig:eva:tpot} report the mean TTFT and TPOT under different request rates. For each model-backend pair and each request rate, we run 1,000 ShareGPT prompts and report the mean latency. Across all evaluated models and request rates, \sys{} consistently achieves lower latency than the baseline, reducing TTFT by 18.8\% and TPOT by 26.9\% on average. The relative improvement is largest under light-to-moderate load, where communication overhead dominates end-to-end latency; at the highest tested request rate, queuing delay becomes a larger fraction of total latency, moderating \sys's relative gain. At 24 requests/s, \sys{} reduces TTFT / TPOT by 15.7\% / 19.0\% on GLM, 14.7\% / 20.5\% on Qwen3-30B-A3B, 17.1\% / 26.2\% on DeepSeek-V2-Lite, and 11.7\% / 22.1\% on GPT-OSS. Notably, on DeepSeek-V2-Lite and GLM-4.7-Flash, the TTFT achieved by \sys{} is close to that of the A800 baseline, and on DeepSeek-V2-Lite, the TPOT is also nearly on par with A800 performance. These results demonstrate the efficiency of the \sys{} technique. By using a host-centric routing mechanism, \sys{} reduces collective communication overhead and improves both first-token latency and steady-state decoding latency.

Notably, GPT-OSS exhibits the largest average TPOT reduction across all request rates. Averaged across all request rates, \sys{} reduces TTFT and TPOT by 17.3\% and 29.9\%, respectively, on GPT-OSS, compared with 21.3\% / 27.0\% on DeepSeek-V2-Lite, 18.7\% / 27.0\% on GLM, and 18.1\% / 23.8\% on Qwen3-30B-A3B. We attribute the strong improvement on GPT-OSS to its model structure. GPT-OSS has a larger hidden dimension (2880) than the other models (2048), which results in higher hidden-state communication volume during MoE routing. Therefore, \sys{}'s host-centric routing mechanism, which reduces collective communication overhead, yields a more pronounced benefit, particularly for decoding latency, on GPT-OSS.

\begin{figure}[t]
\centering
\includegraphics[width=\linewidth]{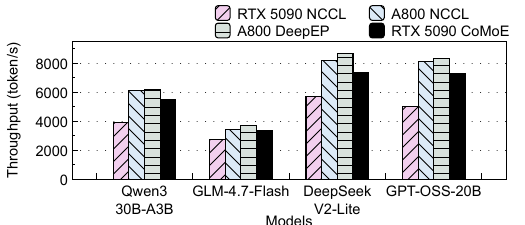}
\caption{Throughput across different models.}
\label{fig:eva:throughput}
\end{figure}

\parabf{Throughput.}
Figure~\ref{fig:eva:throughput} shows the inference throughput of different models, where each model is evaluated on 1000 prompts. 
We primarily compare \sys{} against the default SGLang All-to-All backend with NCCL communication. 
Although DeepEP provides optimized MoE communication primitives, its inference optimization is mainly targeted at newer GPU architectures and large-scale training workloads. 

\sys{} consistently improves throughput over the RTX 5090 baseline across all evaluated models. Specifically, it increases throughput from 3,914.17 to 5,506.00 for Qwen3-30B-A3B (1.41$\times$), from 2,712.19 to 3,367.13 for GLM-4.7-Flash (1.24$\times$), from 5,726.48 to 7,386.28 for DeepSeek-V2-Lite (1.29$\times$), and from 4,991.01 to 7,304.18 for GPT-OSS-20B (1.46$\times$). Notably, \sys{} on RTX 5090 achieves throughput competitive with the A800 baseline, substantially closing the performance gap between consumer and datacenter-grade hardware. Specifically, \sys{} reaches 90.3\% of the A800 throughput for Qwen3-30B-A3B (5,506.00 vs. 6,100.20), 98.8\% for GLM-4.7-Flash (3,367.13 vs. 3,407.00), 90.0\% for DeepSeek-V2-Lite (7,386.28 vs. 8,202.70), and 89.7\% for GPT-OSS-20B (7,304.18 vs. 8,143.10). Particularly, on GLM-4.7-Flash, \sys{} narrows the gap to within 1.2\%, effectively matching server-grade GPU performance, and on the remaining models it stays within about 10\% of the A800 baseline. Even against the stronger A800 DeepEP configuration, \sys{} retains 85.3\%--90.0\% of its throughput. This demonstrates that \sys{} substantially bridges the throughput gap between consumer-grade and server-grade GPUs, making high-performance MoE serving practical on consumer hardware.

These throughput gains are consistent with the latency improvements reported above and further confirm that communication, rather than computation, is the dominant bottleneck on consumer GPU servers. By reducing redundant traffic during dispatch and eliminating unnecessary bulk synchronization during combine, \sys{} shortens the communication critical path and allows GPUs to spend more time on useful expert computation instead of waiting for inter-GPU data exchange to finish. As a result, the system can sustain a substantially higher token processing rate under offline serving workloads.

\parabf{GPU Hardware Cost Estimation.}
To assess the cost efficiency of \sys{}, we conduct a CapEx-based GPU Hardware Cost estimation comparing \sys{} on RTX 5090 against the strongest A800 configuration, SGLang with the DeepEP backend, focusing on the hardware cost required to achieve equivalent serving throughput. As Table~\ref{tab:cost-effectiveness} shows, across all evaluated models, \sys{} on a single RTX 5090 node achieves 85.3\%--90.0\% of the throughput of a single A800 DeepEP node. To match the throughput of an A800 node, only a modest over-provisioning of 1.11$\times$ to 1.17$\times$ RTX 5090 nodes is required. 
Currently, the RTX 5090 is priced at approximately \$1,999 per GPU, making an 8-GPU node cost roughly \$15,992. In contrast, an 8-GPU NVIDIA A800 80GB node commands \$80,000--\$96,000. Factoring in the performance ratio, achieving A800-level throughput with \sys{} requires an effective hardware investment of approximately \$17,760 to \$18,742. Ultimately, \sys{} matches the serving performance of NVLink-capable A800 GPUs running DeepEP at only 18.5\%--23.4\% of the hardware cost.

\begin{table}[t]
  \centering
  \caption{Cost-effectiveness of \sys{} on RTX 5090 versus A800 with DeepEP.}
  \label{tab:cost-effectiveness}
  \setlength{\tabcolsep}{2pt}
  \renewcommand{\arraystretch}{1.15}

  \begin{tabular}{@{}lcccc@{}}
    \toprule
    \textbf{Metric}
      & \shortstack{\textbf{Qwen3}\\\textbf{30B-A3B}}
      & \shortstack{\textbf{GLM-4.7}\\\textbf{Flash}}
      & \shortstack{\textbf{DeepSeek}\\\textbf{V2-Lite}}
      & \shortstack{\textbf{GPT-OSS}\\\textbf{20B}} \\
    \midrule
    \sys{} Thpt.
      & 5,506.00 & 3,367.13 & 7,386.28 & 7,304.18 \\
    A800 Thpt.
      & 6,200.60 & 3,739.40 & 8,656.30 & 8,353.00 \\
    \sys{} Tokens/\$
      & 0.344 & 0.211 & 0.462 & 0.457 \\
    A800 Tokens/\$
      & 0.078 & 0.047 & 0.108 & 0.104 \\
    Tokens/\$ Ratio
      & 4.44$\times$ & 4.50$\times$ & 4.27$\times$ & 4.37$\times$ \\
    \bottomrule
  \end{tabular}
\end{table}

\begin{figure}[t]
\centering
\includegraphics[width=\linewidth]{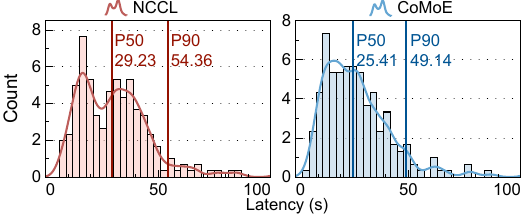}
\caption{PDF and histogram of latencies of DeepResearch tasks.}
\label{fig:evl:deepresearch-pdf}
\end{figure}

\parabf{Real-world Application.} We employ DeepResearch~\cite{tongyidr}—an agent equipped with diverse tool-calling capabilities, including web browsing and text search—to conduct real-world benchmark experiments and evaluate the performance improvements achieved by \sys{}. Specifically, we evaluate \sys{} on the BrowseComp-VL benchmark utilizing the Qwen3-30B-A3B model. We measure the end-to-end latency of each task and compare \sys{} against the baseline. Figure~\ref{fig:evl:deepresearch-pdf} plots the probability density function (PDF) and histogram of the end-to-end latency across all tasks. Although external tool calls diminish the performance gains provided by \sys{}, \sys{} still shifts the latency distribution to the left, reducing the median latency from 29.23s to 25.41s (a 13.1\% reduction) and the \texttt{P90} latency from 54.36s to 49.14s (a 9.6\% reduction). Notably, these gains are consistent across tasks of varying complexity, demonstrating that \sys{} delivers broad latency improvements rather than benefiting only a specific subset of tasks.

%% file: content/6.relatedworks.tex
\section{Related Works}

\subsection{MoE Communication Optimization} 
Communication optimization for MoE models has been extensively studied in prior work. Early systems such as FastMoE~\cite{fastmoe} and FasterMoE~\cite{fastermoe} improve the efficiency of communication during MoE training by smart fine-grained scheduling and computation-communication overlap. Building on these efforts, frameworks such as DeepSpeed-MoE~\cite{deepspeed} and Tutel~\cite{tutel} adopt hierarchical All-to-All to reduce inter-node communication volume, while Aurora~\cite{aurora} further reduces the latency by reordering token transmission. Janus~\cite{janus} introduces an innovative data-centric communication paradigm, which moves experts between GPUs instead of tokens. Meanwhile, specialized communication libraries have also been proposed for MoE workloads. For example, DeepEP~\cite{deepep} and PPLX~\cite{pplx} integrate communication into GPU kernels by leveraging techniques such as NVSHMEM and IBGDA, minimizing synchronization overhead and reducing token dispatch latency by enabling direct data movement within kernels.

However, most existing works focus on high-end GPUs with NVLink support, while communication optimization for MoE inference on consumer GPUs remains underexplored. In contrast, \sys{} provides efficient token dispatch and combine communication support for MoE inference on consumer-grade GPUs, without relying on high-speed point-to-point interconnects such as NVLink or NVSHMEM.

\subsection{Low-cost MoE Inference} 
The astronomical parameter count of modern MoE models poses severe memory challenges, prompting several studies to investigate cost-effective strategies for deploying them on resource-constrained or consumer-grade hardware. To bypass the strict GPU memory wall, previous works like MoE-Infinity~\cite{moe-infinity}, HOBBIT~\cite{hobbit} and ExpertFlow~\cite{expertflow} adopt expert offloading and prefetching strategies to improve MoE inference performance by overlapping experts loading with computation, and FloE~\cite{floe} employs a hybrid compression mechanism on the expert's parameter matrices to achieve inference acceleration on memory-constrained GPUs. Meanwhile, KTransformers~\cite{ktransformers} and Fiddler~\cite{fiddler} target heterogeneous environments, leveraging the computing power and abundant host memory of both CPUs and GPUs to accelerate MoE inference.

Existing low-cost solutions predominantly tackle the single-GPU scenario by focusing on \textit{weight offload} (i.e., fetching expert weights from CPU DRAM to GPU memory on demand) and \textit{computation offload} (CPU-GPU orchestration). In contrast, \sys{} is orthogonal to these studies, as it targets efficient communication in a consumer GPU-only setting.

%% file: content/7.conclusion.tex
\section{Conclusion}

Consumer GPUs deliver datacenter-comparable compute at a fraction of the cost, which are ideally suited for MoE model inference.
However, their limited interconnects leave their full potential untapped. 
We present \sys{}, a communication-efficient MoE inference system tailored for consumer-grade GPUs, which eliminates dispatch redundancy via host-backed token multicast and removes synchronization bottlenecks via a fine-grained host-staged combine. 
Evaluations show \sys{} achieves up to 1.46$\times$ throughput gains over the baselines and reaches 85--90\% of the throughput of NVLink-capable A800 GPUs running DeepEP, matching it at only 23.4\% of the hardware cost.